\documentclass[dvips,12pt,a4paper]{article}
\usepackage{amssymb}
\usepackage{amsmath}
\usepackage{graphicx}
\usepackage{epsfig}

\def\dcoprod{\mathop{\displaystyle \coprod }}%
\def\dprod{\mathop{\displaystyle \prod }}%

\begin{document}

\begin{center}
{{\bf\Large Generalization of the Poisson distribution for the case of changing probabilities of consecutive events}} 
\newline\newline {E.A.Kushnirenko}$^\dagger$
\end{center}

\begin{abstract}

In this paper the generalization of the Poisson distribution is derived for the case when each consecutive event changes event rate. A simple formula for the probability of observing of a given number of events for the selected period of time is derived for a given set of rates. Application of this distribution in high-energy physics calculations is discussed.

\end{abstract}

{\bf Preface}: This distribution was used to perform calculations of the backgrounds in the electron-photon beam interactions~\cite{Kushnirenko1},\cite{Kushnirenko2}.  Mathematical properties of the distribution are discussed here in more detail.  This distribution can be used  when interaction causes change of the detector properties or of the incident beam.  Also it can be of certain interest as an example of yet another generalization of the Poisson distribution.

\section{Introduction}

The Poisson distribution is used to describe a variety of different processes.  It is used very often in the analysis of data from the experimental sets in accelerator physics, in cosmic rays physics, in radioactive decay studies, in fluctuations of energy losses of a particle moving in the matter and in many other cases.  The only parameter that describes the distribution is the rate of events $\mu$ or the average number of events per unit time.  The assumption that this parameter is constant gives the Poisson distribution unique features and a certain beauty, however limits its applications. 

Consider a several hundred $GeV$ electron beam going through a mono\-chromatic photon beam going in the opposite direction.  An electron going through the photon beam, can have several interactions with photons, and the energy of the electron will drop down by an order of magnitude after 2 interactions.  The probability of the interaction of the electron with the mono\-chromatic photon beam of a particular energy is defined by the energy of the electron.  In this situation the rate $\mu$ of the Poisson process, which describes the probability of the electron-photon interaction changes dramatically after each interaction, for this reason even for an estimate of the number of the electron-photon interactions the Poisson distribution can not be used.

Next we consider a process of shooting down a military aircraft with shells. The probability of hitting the aircraft grows after each consecutive hit, as its flying characteristics as well as defense systems go down and it becomes an easier target to hit.  There could be different scenarios which describe variations of $\mu_n$.   Clearly the classical Poison distribution will not describe the process correctly and this problem requires generalization of the Poisson distribution.

Also it is quite natural, that after the interaction the properties of the detector change, which leads to the change of the rate of the detection of events.  For example the detection efficiency can degrade after the interaction of the particle with the detector.

In these examples, the rate of observed events during the observation period varies considerably, while during the period of constant rate $\mu$ the observed process is the Poisson process.  These examples show that it can be useful to find distribution of events in processes where average rate changes.  In this paper we consider a process where the rate $\mu$ changes after each consecutive event is happening, and the process for any length of time between events is Poisson, with different rates $\mu_i$.  

\section{The generalization of the Poisson process}

In general the problem can be formulated as follows: find a probability that for a given period of time ${(0,t)}$ there would be $n$ consecutive events, when the average rate of events $\mu$ is not constant, but changes instantaneously after each consequent event.  Before the first event the rate is $\mu_1$, immediately after the first event and before the second event it is $\mu_2$, immediately after the second event and before third event {\bf $\mu_3$}, immediately after event ${n-1}$ and before event $n$ the rate is $\mu_n$.  In each interval between consecutive events the rate is constant.

Generally speaking $\mu_1, \mu_2, \mu_3, ..., \mu_n, ....$ are independent finite positive numbers.  We will consider that all of these numbers are different unless specifically stated otherwise.

The solution of the problem is the generalization of the Poisson distribution which we look for.  Obviously when ${\mu_1=\mu_2=...= \mu_n=...= \mu}$ the distribution should transform into the Poisson distribution.

We will make the following assumptions: 

\begin{enumerate}

\item \label{assumption_1} The probability ${P_n(t)}$ that during the period of time ${(0,t)}$ there would be ${n\geq1}$ events is defined by the rates ${\mu_1, \mu_2, ... , \mu_{n+1}}$ which characterize the rates before $1, 2, \ldots ,n$-th events and rate before event $n+1$.

\item \label{assumption_2} The probability of an event during the small period of time $\Delta t$ is proportional to the duration $\Delta t$ and the expected rate $\mu_i$ of events.

\item \label{assumption_3} The probability of 2 or more events during the small period of time $\Delta t$ is vanishingly small.

\item \label{assumption_4} For any $n\geq 1$ the event with number $n+1$ takes place later than the event with number $n$.

\end{enumerate}

Using these assumptions the probability $P_0(t+\Delta t)$ that during the period of time $(0,t+\Delta t)$ there are no events means that there are no events neither during the period $(0,t)$, nor during the period $(t,t+\Delta t)$.  For short periods of time $\Delta t$ this probability in accordance with assumptions (\ref{assumption_2}) and (\ref{assumption_3}) is equal to:

\begin{eqnarray}
  P_{0} (t+\Delta t)=P_{0} (t)\cdot (1-\mu _{1} \cdot \Delta t)
\end{eqnarray}

where $P_0(t)$ and $(1-\mu _{1} \cdot \Delta t)$ are probabilities that there are no events neither during the period of time $(0,t)$, nor during the period of time $(t+\Delta t)$.  With $\Delta t\rightarrow 0$ we have a differential equation:
\begin{eqnarray}
  \frac{dP_{0}(t)}{dt}=-\mu_{1} P_{0}(t)
  \label{eq:diff_dP0_dt}
\end{eqnarray}
with the initial condition:
\begin{eqnarray}
  P_{0}(0)=1
\end{eqnarray}
which has the solution 
\begin{eqnarray}
  P_{0}(t)=e^{-\mu_{1}\cdot t}
  \label{eq:P0_t}
\end{eqnarray}

Let us now find out the probability $P_1(t+\Delta t)$ of single event happening during the period of time $(t+\Delta t)$.  In accordance with assumptions (\ref{assumption_1}),(\ref{assumption_2}),(\ref{assumption_3}) there could be 2 possibilities: either the first event happens during the period $(0,t)$ and no events during the period $(t,t+\Delta t)$, or there are no events during the period of time $(0,t)$ and single event during the period $(t+\Delta t)$.  So in accordance with assumptions (\ref{assumption_2}),(\ref{assumption_3}):
\begin{eqnarray}
  P_{1}(t+\Delta t) = P_{1}(t)\cdot (1-\mu _{2}\cdot \Delta t) + P_{0}(t)\cdot \mu _{1} \cdot \Delta t
\end{eqnarray}

Considering the difference $P_1(t+\Delta t) - P_1(t)$ and taking $\Delta t\rightarrow 0$ we obtain differential equation:
\begin{eqnarray}
  \frac{dP_{1}(t)}{dt} = -\mu _{2} P_{1} (t) + \mu _{1} P_{0} (t)
\end{eqnarray}

The solution of this equation using (Eq.~\ref{eq:P0_t}) and initial condition
\begin{eqnarray}
  P_{1} (0)=0
\end{eqnarray}
is given by the formula:
\begin{eqnarray}
  P_{1} (t)=\mu _{1} \left[ \frac{e^{-\mu _{1} t} }{\mu _{2} -\mu _{1} }+\frac{e^{-\mu _{2} t} }{\mu _{1} -\mu _{2} } \right] 
\end{eqnarray}

Analogous to the above one can solve the equation for 2 events with the initial conditions 
\begin{equation}
  P_2(0) = P_3(0) = 0 
\end{equation}
we find that:
\begin{eqnarray}
P_2(t) = \mu_1\mu_2 
\left[ 
    \frac{ e^{-\mu_1 t} }{(\mu_3-\mu_1)(\mu_2-\mu_1)} + \right. \nonumber \\  \left. +
    \frac{ e^{-\mu_2 t} }{(\mu_3-\mu_2)(\mu_1-\mu_2)} + \right. \nonumber \\  \left. +
    \frac{ e^{-\mu_3 t} }{(\mu_2-\mu_3)(\mu_1-\mu_3)} 
\right] 
\end{eqnarray}
\begin{eqnarray}
P_3(t) = \mu_1\mu_2\mu_3 
\left[ 
    \frac{ e^{-\mu_1 t} }{(\mu_4-\mu_1)(\mu_3-\mu_1)(\mu_2-\mu_1)} + \right. \nonumber \\ \left. +
    \frac{ e^{-\mu_2 t} }{(\mu_4-\mu_2)(\mu_3-\mu_2)(\mu_1-\mu_2)} + \right. \nonumber \\ \left. +
    \frac{ e^{-\mu_3 t} }{(\mu_4-\mu_3)(\mu_2-\mu_3)(\mu_1-\mu_3)} + \right. \nonumber \\ \left. +
    \frac{ e^{-\mu_4 t} }{(\mu_3-\mu_4)(\mu_2-\mu_4)(\mu_1-\mu_4)} 
\right] 
\end{eqnarray}

In general case the differential equation is
\begin{eqnarray}
    \frac{dP_{n} (t)}{dt} &=& -\mu_{n+1} P_{n} (t)+\mu _{n} P_{n-1} (t), {\rm when\,\,} n>0 \nonumber \\
    \frac{dP_{n} (t)}{dt} &=& -\mu_{n+1} P_{n} (t)  {\hspace{13ex} \rm when\,\,} n=0
    \label{eq:diff_dPn_dt}
\end{eqnarray}
with initial condition 
\begin{eqnarray}
    P_n(0) &=& 0, {\rm when\,\,} n>0 \nonumber \\
    P_n(0) &=& 1, {\rm when\,\,} n=0
    \label{eq:Pn_eq_0}
\end{eqnarray}

Comparing expressions for $P_0(t), P_1(t), P_2(t), P_3(t)$ and using method of mathematical induction it is straightforward to show that the solution for the probability $P_n(t)$ of $n$ events during the period $(0,t)$ 

\begin{eqnarray}
    P_n(t)=\mu_1\mu_2...\mu_n 
    \left[ 
    \frac { e^{-\mu_1     t} } {(\mu_{n+1}-\mu_1)(\mu_n-\mu_1)...(\mu_2-\mu_1)} + \right. \nonumber \\ \left. +
    \frac { e^{-\mu_2     t} } {(\mu_{n+1}-\mu_2)(\mu_n-\mu_2)...(\mu_1-\mu_2)} + \right. \cdot \cdot \cdot \nonumber \\ \left. \cdot \cdot \cdot +
    \frac { e^{-\mu_{n+1} t} } {(\mu_{n}-\mu_{n+1})(\mu_{n-1}-\mu_{n+1} )...(\mu_1-\mu_{n+1})} 
    \right] 
\end{eqnarray}

The structure of the obtained formula is fairly straightforward: $P_n(t)$ contains $n+1$ terms, where every $j$-th term $(1\leq j \leq n+1)$ contains a fraction with $e^{-\mu_jt}$ in the numerator and the product $(\mu_{n+1}-\mu_j)(\mu_n-\mu_j)\ldots(\mu_1-\mu_j)$ with $n$ factors but without $(\mu_j-\mu_j)$ factor in the denominator.
\begin{figure}
  \includegraphics[width=\textwidth, height=8cm, bb=70 210 540 580]{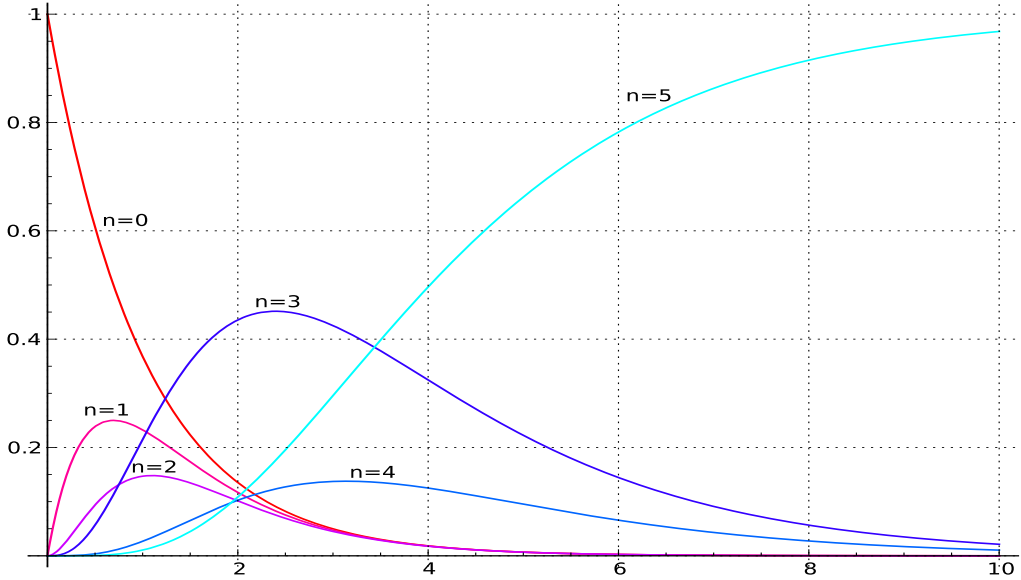}
  \caption {The distributions $P_n(t)$ for $\mu_1=1$,  $\mu_2=2$,  $\mu_3=3$,  $\mu_4=0.5$, $\mu_5=1.50$, and $\mu_6=0$. }
  \label{fig:motivation_dsd0_ratio}
\end{figure}

Expression $P_n(t)$ can be written in more compact form:
\begin{eqnarray}
    P_n(t)=
    \left(\dprod\limits_{i=1}^{n}\mu_i \right) \cdot
    \sum\limits_{i=1}^{n+1}
        \frac{ e^{-\mu_i t} } { \dprod\limits_{j=1;j\neq i}^{n+1} (\mu_j-\mu_i ) }
    \label{poisson_compact}
\end{eqnarray}

Remark: Clearly $P_n(t)$ is a function of not only time $t$, but also of parameters $\mu_1, \mu_2, \ldots, \mu_n, \mu_{n+1}$, so it would be more appropriate to use notation $P_n(t,\mu_1,\mu_2,\ldots,\mu_{n+1})$.  Nevertheless we will use the notation $P_n(t)$ for the sake of shortness, and will use the notation $P_n(t,\mu_1,\mu_2,\ldots,\mu_{n+1})$ in case when it is absolutely necessary.

\section{Analysis of the solution}

\subsection{Expression for the $m$-th derivative of $P_n(t)$ \newline and some other useful formulas} 

Writing down differential equations (Eq.~\ref{eq:diff_dPn_dt}) for $n=1,2,3$ and taking into account (Eq.~\ref{eq:diff_dP0_dt}) we obtain quite helpful for the further calculations system of differential equations:

\begin{eqnarray}
    \frac{dP_0 (t)}{dt} &=&-\mu _1     P_0 (t)                      \nonumber \\
    \frac{dP_1 (t)}{dt} &=&-\mu _2     P_1 (t)+\mu_1 P_0     (t)    \nonumber \\
    \frac{dP_2 (t)}{dt} &=&-\mu _3     P_2 (t)+\mu_2 P_1     (t)    \nonumber \\ \ldots \nonumber \\
    \frac{dP_n (t)}{dt} &=&-\mu _{n+1} P_n (t)+\mu_n P_{n-1} (t) 
    \label{eq:system_diff_dPn_dt}             
\end{eqnarray}

Then
\begin{eqnarray}
    \frac{d P_n(t)}{dt}     &=& (-1)^1 \cdot \mu_{n+1} \cdot P_n(t)       \nonumber \\
                            &+& (-1)^0 \cdot \mu_{n}   \cdot P_{n-1}(t)   \nonumber \\
    \frac{d^2 P_n(t)}{dt^2} &=& (-1)^2 \cdot \mu_{n+1}^2 \cdot P_n(t)     \nonumber \\
                            &+& (-1)^1 \cdot \mu_n \left(\mu_{n+1}+\mu_n\right) \cdot P_{n-1}(t)  \nonumber \\
                            &+& (-1)^0 \cdot \mu_n \mu_{n-1} \cdot P_{n-2}(t)  \nonumber \\
    \frac{d^3 P_n(t)}{dt^3} &=& (-1)^3 \cdot \mu_{n+1}^3 \cdot P_n(t)                             \nonumber \\
                            &+& (-1)^2 \cdot \mu_n \left(\mu_{n+1}^2+\mu_{n+1}\mu_n+\mu_n^2 \right) \cdot P_{n-1}(t)  \nonumber \\
                            &+& (-1)^1 \cdot \mu_n \mu_{n-1} \left(\mu_{n+1}+\mu_{n}+\mu_{n-1} \right) \cdot P_{n-2}(t)  \nonumber \\
                            &+& (-1)^0 \cdot \mu_n \mu_{n-1} \mu_{n-2} \cdot P_{n-3}(t) 
\end{eqnarray}

General formula for the $m$-th derivative of $P_n(t)$, where $(n\geq 0)$, which includes system of equations (Eq.~\ref{eq:system_diff_dPn_dt}) is given by:
\begin{eqnarray}
    \frac{d^m P_n(t)}{dt^m} &=& (-1)^m \mu_{n+1}^m P_n(t)  +  \nonumber \\
                            &+& \sum\limits_{s=1}^m(-1)^{m-s}             \cdot 
    \prod\limits_{j=n-s+1}^{n}\mu_j \cdot 
    \left(\sum\limits_{i=n-s+1}^{n+1}\mu_i\right)^{\otimes (m-s)} \cdot P_{n-s}(t)
    \label{eq:diff_m_dPn_dt} 
\end{eqnarray}
where the following notation is used:\\
\begin{eqnarray}
    \left( \sum\limits_{i=1}^{r}\mu _{i}  \right) ^{\otimes n}
    =\sum\limits_{k_1 + k_2 + \ldots +k_r =n} 
        \mu_1^{k_1} \cdot 
        \mu_2^{k_2} \cdot \ldots \cdot 
        \mu_r^{k_r} ,
    \label{eq:notation_otimes}
\end{eqnarray}
 
For example:
\begin{eqnarray}
    \left( \mu _{1} +\mu _{2} \right) ^{\otimes 3} &=& \mu _{1}^{3} 
    +\mu_{1}^{2} \cdot \mu _{2} +\mu _{1} \cdot \mu _{2}^{2} +\mu _{2}^{3} \nonumber \\
    \left( \mu _{1} +\mu _{2} +\mu_{3} \right) ^{\otimes 2} &=&
    \mu _{1}^{2} +\mu_1\mu_2+\mu_2^2 + \mu_2\mu_3 + \mu_1\mu_3 + \mu_3^2 \nonumber
\end{eqnarray}

Also in usual notation we have:
\begin{eqnarray}
\left( \sum\limits_{i=1}^3\mu_i \right)^{\otimes n} =     
    \sum\limits_{s=0}^n \sum\limits_{m=0}^{s}   \mu_1^{s-m}   \mu_2^{m} \mu_3^{n-s} =
    \sum\limits_{s=0}^n \sum\limits_{m=0}^{n-s} \mu_1^{n-s-m} \mu_2^{m} \mu_3^{s}
\end{eqnarray}
Expression for the derivative (Eq.\ref{eq:diff_m_dPn_dt}) was derived for $n\ge 0$ and $m<n$, but it can be expanded for any $n$ or $m$ if we assume that:
\begin{eqnarray}
\left(\sum\limits_{i=1}^{r}\mu_i \right)^{\otimes n} &=& 1, \mbox{~when~} n=0, \nonumber \\
\label{eq:prodsum_mu}
\left(\sum\limits_{i=1}^{r}\mu_i \right)^{\otimes n} &=& 0, \mbox{~when~} n<0. \\
P_n (t) &=& 0, \mbox{~when~} n<0 \nonumber
\end{eqnarray}

It is interesting to note the behaviour of (Eq.~\ref{eq:diff_m_dPn_dt}) at $t=0$. First of all at $t=0$
\begin{eqnarray}
P_n(0) = 0 \mbox{ for } n\geq 1 \\
P_n(0) = 1 \mbox{ for } n=0
\end{eqnarray}
Then the first term in (Eq.~\ref{eq:diff_m_dPn_dt}) is equal to 0 for $n\geq 1$. In the second term all the terms of the sum vanish except for the term with $s=n$.  According to (Eq.~\ref{eq:prodsum_mu}) we have non-zero terms only when $m\geq n$.  So, given that $P_0(0)=1$, the formula (Eq.~\ref{eq:diff_m_dPn_dt}) at $t=0$ transforms into:
\begin{eqnarray}
  \left. \frac{d^{m} P_{n} (t)}{dt^{m} }\right|_{t=0} = (-1)^{(m-n)} \cdot
  \dcoprod\limits_{j=1}^{n} 
        \mu_j  \cdot \left( 
            \sum\limits_{i=1}^{n+1}\mu_i 
        \right) ^{\otimes (m-n)} \mbox{~when~} m\ge n.
    \label{eq:m-derivative_dP_dt}
\end{eqnarray}
Let us use $m=n+r$ for convenience, then the above formula transforms into
\begin{eqnarray}
\left. \frac{d^{n+r} P_{n} (t)}{dt^{n+r} } \right|_{t=0} = 
    (-1)^{r} \cdot
    \dcoprod\limits_{j=1}^{n}\mu _{j}  \cdot 
        \left(\sum\limits_{i=1}^{n+1}\mu_{i}  \right) ^{\otimes (r)} \mbox{~when~} r\ge 0.
    \label{eq:m-derivative_dP_dt1}
\end{eqnarray}
From expression (Eq.~\ref{eq:diff_m_dPn_dt}) given the remark (Eq.~\ref{eq:prodsum_mu}) we conclude that at $t=0$:
the derivatives : 
\begin{eqnarray}
\left. \frac{d^{m} P_{n} (t)}{dt^{m} } \right| _{t=0} =0, \mbox{~when~} m<n
\end{eqnarray}

According to (Eq.~\ref{eq:m-derivative_dP_dt1}) the first derivative which does not vanish at $t=0$ of the function $P_n(t)$ has the rank $m=n$ and it is equal to 

\begin{eqnarray}
\left. \frac{d^{n} P_{n} (t)}{dt^{n} } \right| _{t=0}
=\dprod\limits_{i=1}^{n}\mu _{i}
\label{eq:dPn_dtn}
\end{eqnarray}

So we can combine the results in the following form:
\begin{eqnarray}
  \left. \frac{d^{m} P_{n} (t)}{dt^{m} }\right| _{t=0} &=& (-1)^{(m-n)} \cdot
  \dcoprod\limits_{j=1}^{n} 
        \mu_j  \cdot \left( 
            \sum\limits_{i=1}^{n+1}\mu_i 
        \right) ^{\otimes (m-n)} \mbox{~when~} m > n. \\
  \left. \frac{d^{m} P_{n} (t)}{dt^{m} } \right| _{t=0} &=&   \dcoprod\limits_{j=1}^{n} 
        \mu_j , \mbox{~when~} m=n \\
  \left. \frac{d^{m} P_{n} (t)}{dt^{m} } \right| _{t=0} &=& 0, \mbox{~when~} m<n
\end{eqnarray}

taking derivative $d^m / dt^m$ in (Eq.~\ref{poisson_compact}) we obtain that 
\begin{eqnarray}
    P_n(t)=
    \left(\dprod\limits_{i=1}^{n}\mu_i \right) \cdot
    \sum\limits_{i=1}^{n+1}
        \frac{ e^{-\mu_i t} } { \dprod\limits_{j=1;j\neq i}^{n+1} (\mu_j-\mu_i ) } \nonumber \\
    \frac{d^{m} P_{n} (t)}{dt^{m} } =
    \left(\dprod\limits_{i=1}^{n}\mu_i \right) \cdot
    \sum\limits_{i=1}^{n+1}
        \frac{ (-1)^m \cdot \mu_i^m \cdot e^{-\mu_i t} } { \dprod\limits_{j=1;j\neq i}^{n+1} (\mu_j-\mu_i ) }
    \label{derivative_poisson_compact}
\end{eqnarray}

Then for any given $m<n$ the sum of the set vanish: 
\begin{eqnarray} 
\sum\limits_{i=1}^{n+1}\frac{(-1)^{m} \cdot \mu _{i}^{m}}
    {\dprod\limits_{j=1;j\neq i}^{n+1}(\mu _{j} -\mu _{i} ) }  =0
\label{eq:sum0_dm_dtm}
\end{eqnarray}

Given the condition $P_n(0)=0$ for $n>0$ the following set vanishes:
\begin{eqnarray}
    \sum\limits_{i=1}^{n+1}
        \frac{1}{\dprod\limits_{j=1;j\neq i}^{n+1}(\mu_j -\mu_i )} = 0, \mbox{~when~} n>0,
\end{eqnarray}
which is a special case of (Eq.~\ref{eq:sum0_dm_dtm}) when $m=0$

Taking the derivative $d^n/dt^n$ of the (Eq.~\ref{poisson_compact}) we conclude that the sum of the following set is equal to 1:
\begin{eqnarray}
    \sum\limits_{i=1}^{n+1}\frac{(-1)^{n} 
        \cdot \mu _{i}^{n}}{\dprod\limits_{j=1;j\neq i}^{n+1}(\mu _{j} -\mu _{i} ) } =1
\end{eqnarray}

In the general case, which applies to any $m$ and $n$ we have:
\begin{eqnarray}
    \sum\limits_{i=1}^{n+1}\frac{(-1)^{m} 
        \cdot \mu _{i}^{m}}{\dprod\limits_{j=1;j\neq i}^{n+1}(\mu _{j} -\mu _{i} ) } = (-1)^{m-n} \cdot 
        \left( 
            \sum \limits_{i=1}^{n+1}\mu_i 
        \right) ^{\otimes (m-n)},
     \label{eq:sum_coeff_zero}
\end{eqnarray}
assuming (Eq.~\ref{eq:prodsum_mu}) for ($m\le n$)

Taking into account (Eq.~\ref{eq:m-derivative_dP_dt}) and (Eq.~\ref{eq:dPn_dtn}) the Taylor series expansion in the vicinity of $t=0$ can be written as:
\begin{eqnarray}
P_{n} (t) 
& = & P_{n}(0) + \sum\limits_{m=1}^{\infty}\frac{t^m}{m!} \cdot \left. \frac{d^{m} P_{n} (t)}{dt^{m} }\right|_{t=0}
  =   P_{n}(0) + \sum\limits_{m=n}^{\infty}\frac{t^m}{m!} \cdot \left. \frac{d^{m} P_{n} (t)}{dt^{m} }\right|_{t=0} \nonumber \\
& = & P_{n}(0) + \sum\limits_{r=0}^{\infty}\frac{t^{n+r}}{(n+r)!} \cdot \left. \frac{d^{n+r} P_{n} (t)}{dt^{n+r} }\right|_{t=0} \nonumber \\
& = & P_{n}(0) + \left( \dcoprod\limits_{j=1}^{n}\mu _{j}  \right) \cdot
      \sum\limits_{r=0}^{\infty }(-1)^{r}  \cdot \frac{t^{n+r} }{(n+r)!} \cdot
      \left( \sum\limits_{i=1}^{n+1}\mu _{i}  \right) ^{\otimes (r)} \nonumber \\
& = & P_{n}(0) + \left( \dcoprod\limits_{j=1}^{n}\mu _{j}  \right) \cdot 
      \frac{t^{n} }{n!} \sum\limits_{r=0}^{\infty }(-1)^{r}  \cdot \left(
      \sum\limits_{i=1}^{n+1}\mu _{i}  \right) ^{\otimes (r)} \cdot \frac{t^{r}
      \cdot n!}{(n+r)!} \nonumber \\ 
\label{eq:Pn_2}
\end{eqnarray}

\subsection{The limiting transition of $P_n(\mu_1,\mu_2,\ldots,\mu_{n+1},t)$ for the case when $(\mu_1,\mu_2,\ldots,\mu_{n+1})\rightarrow \mu$}

For $n>0$ let us consider $P_n(t)$ in the form
\begin{eqnarray}
    P_{n} (t) = \left( \dprod\limits_{j=1}^{n}\mu _{j}  \right) \cdot
        \frac{t^{n} }{n!} \cdot 
        \left[ 
            \sum\limits_{r=0}^{\infty }(-1)^{r} \frac{n!}{(n+r)!}  \cdot 
            t^{r} \cdot 
            \left( \sum\limits_{i=1}^{n+1}\mu_{i} \right) ^{\otimes r} 
        \right]
    \label{eq:pn_close_mu}
\end{eqnarray}

Let all $\mu_i = \mu$. According to the notation (Eq.~\ref{eq:notation_otimes}):

\begin{eqnarray}
\left( \sum\limits_{i=1}^{n+1}\mu _{i}  \right) ^{\otimes r} = 
    \sum\limits_{k_1+k_2+\ldots+k_{n+1}=r}^{}
        \mu_{1}^{k_1 } \cdot   
        \mu_{2}^{k_2}  \cdot \ldots \cdot 
        \mu_{n+1}^{k_{n+1} }
    \label{eq:sum_mi_times_r}
\end{eqnarray}

The number $N$ of terms in the expression (Eq.~\ref{eq:sum_mi_times_r}) is given by the formula:
\begin{eqnarray}
N=C_{n+r}^{r} =\frac{(n+r)!}{n!\cdot r!}.
\end{eqnarray}
Hence using formula (Eq.~\ref{eq:pn_close_mu})
\begin{eqnarray}
    \lim_{\mu _{i} \rightarrow \mu} P_{n} (t) 
        &=& \frac{\mu ^{n} t^{n} }{n!} 
            \sum\limits_{r=0}^{\infty }(-1)^{r}  
            \cdot \frac{t^r}{r!} \cdot \mu^r \nonumber \\
        &=& \frac{\left( \mu \cdot t\right) ^{n} }{n!} \cdot e^{-\mu \cdot t}, \mbox{~when~} n>1 \\
    \lim_{\mu _{1} \rightarrow \mu} P_{0} (t) 
        &=&  e^{-\mu t}, \mbox{~when~} n=0
\end{eqnarray}

So for any $n\ge 0$ generalized Poisson distribution transforms into Poisson distribution when $(\mu_1,\mu_2,\ldots,\mu_{n+1})\rightarrow \mu$

\subsection{Case of $\mu_{n+1}=0$}

It is quite clear that if $\mu_{n+1}=0$, which basically means that there are only $n$ events, and event $n+1$ would never happen, then $P_n(t)$ is assimptotically growing to 1, and $P_{n+1}(t) = P_{n+2}(t) = \ldots = 0$.  Let us consider specific cases:

$\mu_1>0,\mu_2=0$, then
\begin{equation}
  P_1(t)=\mu_1 \left[ \frac{ e^{-\mu_1 t }} {\mu_2-\mu_1} + \frac{e^{-\mu_2 t}}{\mu_1-\mu_2} \right] = 1 - e^{-\mu_1 t} = 1 - P_0(t)
\end{equation}

$\mu_{1,2}>0,\mu_3=0$, then
\begin{eqnarray}
P_2(t) &=& \mu_1\mu_2 
\left[ 
    \frac{ e^{-\mu_1 t} }{(-\mu_1)(\mu_2-\mu_1)} + \frac{ e^{-\mu_2 t} }{(-\mu_2)(\mu_1-\mu_2)} + \frac{ 1 }{\mu_2\cdot\mu_1} 
\right] \nonumber \\
       &=& 1 - \mu_1 \frac{ e^{-\mu_2 t} }{\mu_1-\mu_2} - \mu_2 \frac{ e^{-\mu_1 t} }{\mu_2-\mu_1} \nonumber \\
       &=& 1 - P_1(t) + \mu_1 \frac{ e^{-\mu_1 t} }{\mu_2-\mu_1} - \mu_2 \frac{ e^{-\mu_1 t} }{\mu_2-\mu_1} \nonumber \\
       &=& 1 - P_1(t) - P_0(t)
\end{eqnarray}

It is straightforward to show than when $\mu_{1,\ldots, n-1}>0,\mu_n=0$ then
\begin{eqnarray}
    P_{n-1}(t)&=&\dprod\limits_{i=1}^{n-1}\mu_i \cdot
    \sum\limits_{i=1}^{n}
        \frac{ e^{-\mu_i t} } { \dprod\limits_{j=1;j\neq i}^{n} (\mu_j-\mu_i ) } \nonumber \\
    &=& 1 + \dprod\limits_{i=1}^{n-1}\mu_i \cdot
    \sum\limits_{i=1}^{n-1}
        \frac{ e^{-\mu_i t} } { (-\mu_i)\cdot\dprod\limits_{j=1;j\neq i}^{n-1} (\mu_j-\mu_i ) } \nonumber \\
    &=& 1 - P_{n-2}(t) - \ldots - P_1(t) - P_0(t)
    \label{eq:p_n-1_when0}
\end{eqnarray}

Indeed assuming (Eq.~\ref{eq:p_n-1_when0}) being true, we can derive that, when 
\mbox{$\mu_{1,\ldots,n}>0$}, and \mbox{$\mu_{n+1}=0$} then
\begin{eqnarray}
    P_{n}(t)&=&\dprod\limits_{i=1}^{n}\mu_i \cdot
    \sum\limits_{i=1}^{n+1}
        \frac{ e^{-\mu_i t} } { \dprod\limits_{j=1;j\neq i}^{n+1} (\mu_j-\mu_i ) } \nonumber \\
    &=& 1 + \dprod\limits_{i=1}^{n}\mu_i \cdot
    \sum\limits_{i=1}^{n}
        \frac{ e^{-\mu_i t} } { (-\mu_i)\cdot\dprod\limits_{j=1;j\neq i}^{n} (\mu_j-\mu_i ) } \nonumber \\
    &=& 1 - \dprod\limits_{i=1}^{n-1}\mu_i \cdot
        \frac{ e^{-\mu_n t} } { \dprod\limits_{j=1;j\neq i}^{n} (\mu_j-\mu_i ) } + 
    \dprod\limits_{i=1}^{n}\mu_i \cdot \sum\limits_{i=1}^{n-1}
        \frac{ e^{-\mu_i t} } { (-\mu_i)\cdot\dprod\limits_{j=1;j\neq i}^{n} (\mu_j-\mu_i ) } \nonumber \\
    &=& 1 - P_{n-1}(t) 
    + \dprod\limits_{i=1}^{n-1}\mu_i \cdot \sum\limits_{i=1}^{n-1}
        \frac{ e^{-\mu_i t} } {\dprod\limits_{j=1;j\neq i}^{n} (\mu_j-\mu_i ) } 
    + \dprod\limits_{i=1}^{n-1}\mu_i \cdot \sum\limits_{i=1}^{n-1}
        \frac{ \mu_n \cdot e^{-\mu_i t} } { (-\mu_i)\cdot\dprod\limits_{j=1;j\neq i}^{n} (\mu_j-\mu_i ) } \nonumber \\
    &=& 1 - P_{n-1}(t)  + \dprod\limits_{i=1}^{n-1}\mu_i \cdot \sum\limits_{i=1}^{n-1}
        \frac{ e^{-\mu_i t} } { \dprod\limits_{j=1;j\neq i}^{n} (\mu_j-\mu_i ) } \left[1+ \frac{\mu_n}{-\mu_i}\right] \nonumber \\
    &=& 1 - P_{n-1}(t)  + \dprod\limits_{i=1}^{n-1}\mu_i \cdot \sum\limits_{i=1}^{n-1}
        \frac{ e^{-\mu_i t} } { (-\mu_i) \dprod\limits_{j=1;j\neq i}^{n-1} (\mu_j-\mu_i ) } {\mathrm{\;\;hence\,\,using\,(Eq.\ref{eq:p_n-1_when0})}} \nonumber \\
    &=& 1 - P_{n-1}(t) - P_{n-2}(t) - \ldots - P_1(t) - P_0(t)
    \label{eq:p_n_when0}
\end{eqnarray}

\subsection{The Integral of $P_n(t)$    }

There are several ways to prove that:

\begin{eqnarray} 
  \mu _{n+1} \cdot \int\limits_{0}^{\infty }P_{n} (t)\cdot dt=1 
  \label{eq:integral_P_n}
\end{eqnarray}

Indeed using direct integration and (Eq.~\ref{eq:p_n_when0})
\begin{eqnarray} 
  \int\limits_{0}^{\infty }P_{n} (t)\cdot dt &=& \dprod\limits_{i=1}^n \mu_i \cdot \sum\limits_{i=1}^{n+1}
        \frac{ 1 } { \mu_i \dprod\limits_{j=1;j\neq i}^{n+1} (\mu_j-\mu_i ) }  
    = \frac{1}{\mu_{n+1}} \dprod\limits_{i=1}^{n+1} \mu_i \cdot \sum\limits_{i=1}^{n+1}
        \frac{ 1 } { \mu_i  \dprod\limits_{j=1;j\neq i}^{n+1} (\mu_j-\mu_i ) }  \nonumber \\
    &=& \frac{1}{\mu_{n+1}}\left[P_{n}(0) + P_{n-1}(0) + \ldots + P_1(0) + P_0(0)\right] = \frac{1}{\mu_{n+1}}
\end{eqnarray}

Similarly as $ \mu_1\int\limits_{0}^{\infty }P_{0} (t)\cdot dt=1 $ then using mathematical induction applied to (Eq.~\ref{eq:diff_dPn_dt}) one can prove that
\begin{eqnarray} 
  \frac{dP_{n} (t)}{dt} &=& -\mu_{n+1} P_{n} (t)+\mu _{n} P_{n-1} (t) \nonumber \\
  \mu_{n+1}\int\limits_{0}^{\infty }P_{n} (t)\cdot dt &=&  P_n(0) - P_n(\infty) + \mu_n \cdot \int\limits_{0}^{\infty }P_{n-1} (t)\cdot dt = 1
\end{eqnarray}

\subsection{Normalisation of $P_n(t)$}

Let us prove that the limit of the sum $S(t)=\sum\limits_{n=0}^\infty P_n(t)$ is equal to 1, at any given time $t$. 

In the special case when one of the $\mu_i=0$ the formula (Eq.~\ref{eq:p_n_when0}) can be applied:
\begin{eqnarray}
  \sum\limits_{i=0}^{n-1}P_{i} (t) &=& 1 - P_n(t)  \nonumber \\
  P_{n+1} &=& P_{n+2} = \ldots = 0,  \nonumber
\end{eqnarray} then clearly $\sum\limits_{n=0}^\infty P_n(t)=1$

For the case when every $\mu_i> 0$ let us consider the sum:
\begin{eqnarray}
S_{n} (t)=\sum\limits_{i=0}^{n}P_{i} (t)
\end{eqnarray}

Let us find the limit, to which the first derivative $S_n(t)$ is approximating when $n$ is growing. 
\begin{eqnarray}
  \lim_{n\rightarrow \infty} \frac{dS_{n} (t)}{dt}=\sum\limits_{i=0}^{\infty }\frac{dP_{i} (t)}{dt}
  \label{eq:limit_dS_dt}
\end{eqnarray}

Considering the system of equations (Eq.~\ref{eq:system_diff_dPn_dt}) and (Eq.~\ref{eq:integral_P_n})
\begin{eqnarray}
  \sum \limits_{i=0}^{n} \frac{dP_{i}(t)}{dt} = -\mu_{n+1} \cdot P_{n}(t) = -\frac{\int\limits_{0}^{t}P_{n} (t)\cdot dt}
    {\int\limits_{0}^{\infty }P_{n} (t)\cdot dt}
  \label{eq:sum_dP_dt}
\end{eqnarray}

Clearly if we fix $t$ and keep increasing $n$ the fraction in (Eq.~\ref{eq:sum_dP_dt}) vanishes.  So for any given $t\geq 0$ and for any given set 
$\mu_i>0$:

\begin{eqnarray}
  \frac{dS(t)}{dt} = \lim_{n\rightarrow \infty } \frac{dS_{n} (t)}{dt} =0
\end{eqnarray}
and hence $S(t) = S(0) = \mbox{const}$

So, taking into account (Eq.~\ref{eq:Pn_eq_0}) we derive that $S_n(0)=1$, for $n>0$. That means that for any $t\geq 0$ and for any given set $\mu_i$ the normalization of $P_n(t)$ is:
\begin{eqnarray}
  \sum\limits_{i=0}^{\infty}P_{i} (t) = S(t)= S(0) = 1
\end{eqnarray}

\subsection{The symmetry of $P_n(\mu_1,\mu_2,\ldots,\mu_{n+1},t)$ under the \\ permutation of parameters $\mu_i,\mu_j$}

One of very obvious properties of functions $P_n(t)$ is its independence from mutual interchange of parameters $\mu_1, \mu_2, \ldots, \mu_n$, because of terms simply interchange in (Eq.~\ref{poisson_compact}). 
\begin{eqnarray}
    P_n(t)=
    \left(\dprod\limits_{i=1}^{n}\mu_i \right) \cdot
    \sum\limits_{i=1}^{n+1}
        \frac{ e^{-\mu_i t} } { \dprod\limits_{k=1;k\neq i}^{n+1} (\mu_k-\mu_i ) } 
    \nonumber 
\end{eqnarray}

This gives us the possibility to put parameters $\mu_i$ in any order, for example in the order of growing $\mu_i$.

Let us remark, that this does not apply to parameter $\mu_{n+1}$.  In this case when we interchange $\mu_i$ with $\mu_{n+1}$, the distribution changes according to:
$P_n(\mu_1,\ldots,\mu_i,\ldots,\mu_n,\mu_{n+1},t)\cdot\mu_{n+1} = P_n(\mu_1,\ldots,\mu_{n+1},\ldots,\mu_n,\mu_i,t)\cdot\mu_i$

So the distribution changes under the permutation of parameters $\mu_i,\mu_j$ according to:
\begin{eqnarray}
P_n(\mu_1,\ldots,\mu_i,\mu_j,\ldots,\mu_n,\mu_{n+1},t) &=& P_n(\mu_1,\ldots,\mu_j,\mu_i,\ldots,\mu_n,\mu_{n+1},t)
\label{eq:permutation-1} \\
P_n(\mu_1,\ldots,\mu_i,\ldots,\mu_n,\mu_{n+1},t) &=& P_n(\mu_1,\ldots,\mu_{n+1},\ldots,\mu_n,\mu_i,t)\frac{\mu_i}{\mu_{n+1}}
\label{eq:permutation-2}
\end{eqnarray}

Given this property when studying properties of function $P_n(t)$ it is usually enough to study dependency $P_n(\mu_1,\mu_2,\ldots ,\mu_n,\mu_{n+1},t)$ with respect to $\mu_{n+1}$ and with respect to any of the parameters $\mu_1,\ldots , \mu_n$, and we can use $\mu_1$ for example.

\subsection{The limiting transition in $P_n(\mu_1,\mu_2,\ldots,\mu_{n+1},t)$ when several parameters $\mu_i$ are equal to each other}.

Let us assume that $k$ different parameters $\mu_i$ are equal to each other.  Due to the symmetry of $P_n$ under the permutation of parameters $\mu_i$ we can calculate the case of $P_n(\mu_1=\mu_2=\ldots=\mu_k=\mu,\mu_{k+1},\ldots,\mu_{n+1})$  

Let us consider $k=2$, then $\mu_1=\mu$ and we calculate the limit when $\mu_2\rightarrow\mu$:

\begin{eqnarray}
\frac{P_n}{\dprod\limits_{i=1}^{n}\mu_i} 
 &=& \frac{e^{-\mu\cdot t}}{(\mu_2-\mu)\ldots(\mu_{n+1}-\mu)} 
 + \frac{e^{-\mu_2\cdot t}}{(\mu-\mu_2)\ldots(\mu_{n+1}-\mu_2)}
 + \sum\limits_{i=3}^{n+1}\frac{e^{-\mu_i\cdot t}}{\dprod\limits_{j=1;j\neq i}^{n+1} (\mu_j-\mu_i)} \nonumber \\
\end{eqnarray}

Let us define the third term as $R(3)$ and define $\Delta\mu = \mu_2-\mu$, then expanding in terms power of $\Delta\mu$

\begin{eqnarray}
\frac{P_n}{\dprod\limits_{i=1}^{n}\mu_i} 
 &=& \frac{e^{-\mu\cdot t}}{\Delta\mu(\mu_3-\mu)\ldots(\mu_{n+1}-\mu)} \left[ 1
 + \frac{e^{         -\Delta\mu\cdot t}}{(-1)
            (1-\frac{\Delta\mu}{\mu_3-\mu})\ldots (1-\frac{\Delta\mu}{\mu_{n+1}-\mu} )  } 
 \right] + R(3) \nonumber \\
 &=& 
\frac{e^{-\mu\cdot t}}{\Delta\mu(\mu_3-\mu)\ldots(\mu_{n+1}-\mu)} \left [ 1 
- \left(1 + (-\Delta\mu\cdot t) + \frac{(-\Delta\mu\cdot t)^2}{2!} + \ldots \right)\times  \right.
 \nonumber \\
 &\times&
   \left. 
      \dprod\limits_{i=3}^{n+1}\left(1 + \frac{\Delta\mu}{\mu_i-\mu} + \left(\frac{\Delta\mu}{\mu_i-\mu}\right)^2 +\ldots \right)    
   \right] + R(3) \nonumber \\
 &=&
\frac{e^{-\mu\cdot t}}{\Delta\mu(\mu_3-\mu)\ldots(\mu_{n+1}-\mu)} \left [ 1 
- \left(1 + \left[ \sum_{i=3}^{n+1}\frac{\Delta\mu}{\mu_i-\mu} -\Delta\mu\cdot t \right] + \right.\right.
 \nonumber \\
 &+&
   \left. \left.  \left[\frac{(-\Delta\mu\cdot t)^2}{2!} 
        + (-\Delta\mu\cdot t)\left(\sum_{i=3}^{n+1}\frac{\Delta\mu}{\mu_i-\mu}\right)^{\otimes 1} 
        + \left(\sum_{i=3}^{n+1}\frac{\Delta\mu}{\mu_i-\mu}\right)^{\otimes 2} 
    \right] 
  + \ldots \right.\right. \nonumber \\
 &+&
  \left.\left. \sum_{k=0}^{n}\left[\frac{(-\Delta\mu\cdot t)^k}{k!} 
               \left(\sum_{i=3}^{n+1}\frac{\Delta\mu}{\mu_i-\mu}\right)^{\otimes {n-k}} \right] + \ldots
  \right)  \right] + R(3)  \\
\end{eqnarray}
where notation (Eq.~\ref{eq:notation_otimes}) was used.  In the limit $\Delta\mu\rightarrow 0$

\begin{eqnarray}
\frac{P_n(\mu_1,\mu_2=\mu)}{\dprod\limits_{i=1}^{n}\mu_i} 
  &=& \frac{e^{-\mu\cdot t}}{\dprod\limits_{i=3}^{n+1}(\mu_i-\mu)} 
        \left[ t + \sum_{i=3}^{n+1}\frac{-1}{\mu_i-\mu}
        \right] + R(3) \nonumber \\
\frac{P_n(\mu_1,\mu_2,\mu_3=\mu)}{\dprod\limits_{i=1}^{n}\mu_i} 
  &=& \frac{e^{-\mu\cdot t}}{\dprod\limits_{i=4}^{n+1}(\mu_i-\mu)} 
        \left[ \frac{t^2}{2!}   + t\cdot\sum_{i=4}^{n+1}\frac{-1}{\mu_i-\mu}
                                + \left(\cdot\sum_{i=4}^{n+1}\frac{-1}{\mu_i-\mu}\right)^{\otimes 2}
        \right] + R(4) \nonumber \\
\frac{P_n(\mu_1,\mu_2,\ldots,\mu_k=\mu)}{\dprod\limits_{i=1}^{n}\mu_i} 
  &=& \frac{e^{-\mu\cdot t}}{\dprod\limits_{i=k+1}^{n+1}(\mu_i-\mu)} 
        \left[ \sum_{j=0}^{k-1} 
            \frac{t^{k-j}}{(k-j)!} \times 
            \left(\sum_{i=k+1}^{n+1}\frac{-1}{\mu_i-\mu}\right)^{\otimes j}
        \right] + R(k+1) \nonumber 
\end{eqnarray}
where we use the notation
\begin{equation}
 R(k) = \sum\limits_{i=k}^{n+1}\frac{e^{-\mu_i\cdot t}}{\dprod\limits_{j=1;j\neq i}^{n+1} (\mu_j-\mu_i)} 
\end{equation}

In a particular case when only $\mu_{n+1}$ differs from the rest of $\mu_i = \mu$ we obtain:
\begin{eqnarray}
P_2(\mu_1,\mu_2=\mu,\mu_3) &=& \mu^2 
    \left( \frac{e^{-\mu_3 \cdot t}}{(\mu -\mu_3)^2} - e^{-\mu\cdot t} 
        \left[  \frac{1}{(\mu -\mu _{3} )^{2} } +\frac{t}{(\mu -\mu _{3} )} \right]  
    \right) \nonumber \\
P_3(\mu_1,\mu_2,\mu_3=\mu,\mu_4) &=& \mu^3 
    \left( \frac{e^{-\mu_4 \cdot t}}{(\mu -\mu_4)^3} - e^{-\mu\cdot t} 
        \left[ \frac{1}{(\mu -\mu _{4} )^{3} } +\frac{t}{(\mu
-\mu _{4} )^{2} } +\frac{t^{2} }{2!(\mu -\mu _{4} )} \right]  
    \right) \nonumber \\
P_n(\mu_1,\ldots,\mu_n=\mu,\mu_{n+1}) &=& \mu^n 
    \left( \frac{e^{-\mu_{n+1} \cdot t}}{(\mu -\mu_{n+1})^n} - e^{-\mu\cdot t} 
        \left[ \sum\limits_{j=0}^{n-1}\frac{t^j}{j!\left( \mu -\mu _{n+1} \right) ^{(n-j)} } \right]  
    \right)
\end{eqnarray}

\subsection{The limiting transition in $P_n(\mu_1,\mu_2,\ldots,\mu_{n+1},t)$ for the case when $\mu_i\rightarrow \infty$}

It is easy to see from compact form of the distribution (Eq.~\ref{poisson_compact}):
\begin{eqnarray}
    P_n(t)=
    \left(\dprod\limits_{i=1}^{n}\mu_i \right) \cdot
    \sum\limits_{i=1}^{n+1}
        \frac{ e^{-\mu_i t} } { \dprod\limits_{j=1;j\neq i}^{n+1} (\mu_j-\mu_i ) }
    \nonumber
\end{eqnarray}
that

\begin{eqnarray}
\lim_{\mu_1\rightarrow \infty } P_{n} (\mu_1,\mu_2,\ldots,\mu_{n+1},t) &=& 
\left(\dprod\limits_{i=2}^{n}\mu _{i}  \right) \cdot
\sum\limits_{i=2}^{n+1}\frac{e^{-\mu_i\cdot t}}{\dprod\limits_{j=2,j\neq i}^{n+1}(\mu_j-\mu_i)} \,\rm{, or}\nonumber \\
\lim_{\mu_1\rightarrow \infty } P_{n} (\mu_1,\mu_2,\ldots,\mu_{n+1},t) &=& 
P_{n-1} (\mu_2,\ldots,\mu_{n+1},t) \nonumber
\end{eqnarray}
which certainly makes sense, as in the limit $\mu_1\rightarrow\infty$ 1-st event happens right away and following events obey $P_{n-1}(t)$ distribution. Due to the symmetry under the permutations (Eq.~\ref{eq:permutation-1}) same formula applies to any $\mu_i$ where $i\leq n$.

When the last parameter $\mu_{n+1}\rightarrow\infty$, then using (Eq.~\ref{eq:permutation-2}) 

\begin{eqnarray}
\lim_{\mu_{n+1}\rightarrow \infty} P_{n} (\mu_1,\ldots,\mu_n,\mu_{n+1},t) &=&
\lim_{\mu_{n+1}\rightarrow \infty} P_{n} (\mu_1,\ldots,\mu_{n+1},\mu_n,t) \frac{\mu_n}{\mu_{n+1}}
\nonumber \\
&=& P_{n-1} (\mu_1,\ldots,\mu_n,t) \frac{\mu_n}{\mu_{n+1}} 
\nonumber
\end{eqnarray}
 we can unify the result in the following way:

\begin{eqnarray}
\lim_{\mu_i\rightarrow \infty} P_{n}(t)&=&P_{n-1} (\mu_1,\ldots,\mu_{i-1},\mu_{i+1},\ldots,\mu_{n+1},t)\,\, {\rm when}\, i\leq n \\
\lim_{\mu_i\rightarrow \infty} P_{n}(t)&=&P_{n-1} (\mu_1,\ldots,\mu_{n},t) \frac{\mu_n}{\mu_{n+1}}\,\, {\rm when}\, i=n+1
\end{eqnarray}

\subsection{Partial derivative $\partial P_{n}/\partial \mu_i $}

Let us first calculate $\partial P_{n}/\partial \mu_{n+1}$ and for any other $\mu_i$ the partial derivative can be calculated using the symmetry (Eq.~\ref{eq:permutation-1}) and (Eq.~\ref{eq:permutation-2}) under the permutation of the parameters $\mu_i$.

\begin{eqnarray}
\frac{\partial P_{1} (\mu _{1} ,\mu _{2} ,t)}{\partial \mu _{2} } &=&-\mu
_{1} \exp (-\mu _{2} t)\cdot 
    \left[ \frac{e^{-(\mu _{1} -\mu _{2} )t} +(\mu _{1} -\mu _{2} )t-1}{(\mu _{2} -\mu _{1} )^{2} } \right] \nonumber \\
    \frac{\partial P_{2} (\mu _{1} ,\mu _{2} ,\mu _{3} ,t)}{\partial \mu _{3}} 
    &=&-\mu _{1} \mu _{2} e^{-\mu _{3} t} \cdot \left[ \frac{e^{-(\mu _{1}-\mu _{3} )t}  +(\mu _{1} -\mu _{3} )t-1}{(\mu _{3} -\mu _{1} )^{2} (\mu
_{2} -\mu _{1} )} \right. \nonumber \\
  &+& \left.\frac{e^{-(\mu _{2} -\mu _{3} )t} +(\mu _{2} -\mu _{3}
)t-1}{(\mu _{3} -\mu _{2} )^{2} (\mu _{1} -\mu _{2} )} \right] \nonumber \\
    \frac{\partial P_{3} (\mu _{1} ,...,\mu _{4} ,t)}{\partial \mu _{4} }
 &=&-\left( \dprod\limits_{i=1}^{3}\mu _{i}  \right) e^{-\mu _{4} t} \left[
        \frac{e^{-(\mu _{1} -\mu _{4} )t} +(\mu _{1} -\mu _{4} )t-1}{(\mu _{4}-\mu _{1} )^{2} (\mu _{3} -\mu _{1} )(\mu _{2} -\mu _{1} )} \right. \nonumber \\
 &+&\left.\frac{e^{-(\mu _{2} -\mu _{4} )t} +(\mu _{2} -\mu _{4} )t-1}{(\mu _{4}
-\mu _{2} )^{2} (\mu _{3} -\mu _{2} )(\mu _{1} -\mu _{2} )} +\right.  \nonumber \\
  &+&\left. \frac{e^{-(\mu _{3} -\mu _{4} )t} +(\mu _{3} -\mu _{4} )t-1}{(\mu
_{4} -\mu _{3} )^{2} (\mu _{2} -\mu _{3} )(\mu _{1} -\mu _{3} )} \right] \nonumber \\
\frac{\partial P_{n} (\mu_1,\ldots,\mu _{n+1},t)}{\partial \mu_{n+1} }
&=&-\left( \dprod\limits_{i=1}^{n}\mu _{i}  \right) 
\cdot e^{-\mu _{n+1} t} \sum\limits_{i=1}^{n}\frac{e^{-(\mu_{i} -\mu _{n+1})t} +(\mu _{i}-\mu _{n+1})t-1}{(\mu _{n+1} -\mu _{i} )^{2} \cdot
\dprod\limits_{j=1,j\neq i}^{n}(\mu _{j} -\mu _{i} ) }
\end{eqnarray}

It is straightforward to show that $\partial P_{n}/\partial \mu_{n+1}<0 $.  Indeed expanding $e^{-(\mu_{i} -\mu _{n+1})t}$ we obtain:
\begin{eqnarray}
\frac{\partial P_{n} (\mu_1,\ldots,\mu _{n+1},t)}{\partial \mu_{n+1} } &=& - \dprod\limits_{i=1}^{n}\mu _{i} 
\cdot e^{-\mu _{n+1} t} \sum\limits_{i=1}^{n}\frac{\sum\limits_{k=2}^{\infty} (-1)^k(\mu_i -\mu _{n+1} )^{k} \frac{\cdot t^k}{k!}}{(\mu _{n+1} -\mu _{i} )^{2} \cdot
\dprod\limits_{j=1,j\neq i}^{n}(\mu _{j} -\mu _{i} ) } \nonumber \\
&=& 
- \dprod\limits_{i=1}^{n}\mu _{i} 
\cdot e^{-\mu _{n+1} t} \sum\limits_{k=0}^{\infty}\frac{t^{k+2}}{(k+2)!}\sum\limits_{i=1}^{n}\frac{ (-1)^k(\mu_i -\mu _{n+1} )^{k} }
{\dprod\limits_{j=1,j\neq i}^{n}(\mu _{j} -\mu _{i} ) } \nonumber \\
\end{eqnarray}

Then using (Eq.~\ref{eq:sum_coeff_zero})
\begin{eqnarray}
\frac{\partial P_{n} (\mu_1,\ldots,\mu _{n+1},t)}{\partial \mu_{n+1} }
 &=& - \dprod\limits_{i=1}^{n}\mu _{i} 
\cdot e^{-\mu _{n+1} t} \sum\limits_{k=0}^{\infty}\frac{t^{k+2}}{(k+2)!} (-1)^{k-n+1}\left(\sum\limits_{i=1}^{n} \mu_i -\mu _{n+1} \right)^{\otimes{k-n+1}} \nonumber \\ 
&=& - \dprod\limits_{i=1}^{n}\mu _{i} 
\cdot e^{-\mu _{n+1} t} \sum\limits_{r=0}^{\infty}\frac{t^{n+r+1}}{(n+r+1)!} (-1)^{r}\left(\sum\limits_{i=1}^{n} \mu_i -\mu _{n+1} \right)^{\otimes{r}} \nonumber \\ 
\end{eqnarray}

The last expression is similar to the Taylor series expansion of \mbox{$P_{n+1}(\mu_i-\mu_{n+1},t)$} (Eq.~\ref{eq:Pn_2}) which is always positive.  Hence  $\partial P_{n}/\partial \mu_{n+1}<0 $ 
\subsection{Analysis of functions $P_n(t)$}

Expression for $P_n(t)$ where $(n\geq 1)$ (Eq.~\ref{poisson_compact}) is the sum of monotonically decreasing functions or monotonically increasing functions, depending on the sign of coefficeint before exponent.  It is clear that $P_n(t)$ is unique, smooth, differentiable function with no discontinuities, which is equal to 0 when $t=0$ and when $t\rightarrow \infty$.  From the Taylor series expansion (Eq.~\ref{eq:Pn_2})  the function $P_n(t)$ obeys the following inequality:

\begin{eqnarray}
\left(\dprod\limits_{i=1}^n\mu_i \right) \cdot \frac{t^{n}}{n!} \cdot \exp{\left[ -\sum\limits_{i=1}^{n+1}\mu_i \cdot t\right]}
    < P_{n} (t) <
\left(\dprod\limits_{i=1}^n\mu_i \right) \cdot \frac{t^{n}}{n!} \cdot e^ {-\mu _{\min } \cdot t}
\end{eqnarray}
where $\mu_{\min}=\min(\mu_1, \mu_2, \ldots , \mu_{n+1}$).  And it quite clear that $P_n(t)$ is positive, bounded function.

\section{Conclusion}

So in this paper the generalization of the Poisson distribution is found for the case of changing rates caused by consecutive events.  In case of constant event rate the distribution naturally transforms in the classical Poisson distribution.  The derived generalization can have different applications, especially in simulation of cascade processes.  It is possible that this problem was already solved and published, but the author did not manage to find such a publication.  The author is grateful to S.M.Sergeev for the usefull discussion.

\end{document}